\documentclass{aa} 
\usepackage{graphicx}
\usepackage{txfonts}
\usepackage{natbib}
\bibpunct{(}{)}{;}{a}{}{,}
\usepackage[colorlinks=true,linkcolor=blue,citecolor=blue]{hyperref}

\DeclareTextSymbol{\degre}{OT1}{23}

\begin{document}

\title{A rare sextuple-merging brightest cluster galaxy system in a disturbed galaxy cluster observed with the
\textit{Einstein} Probe Follow-up X-ray Telescope}
\titlerunning{A rare merging BCG}

\author{Z. L. Wen\inst{1,2,3}\corrauth{zhonglue@nao.cas.cn}
  \and S. M. Jia\inst{3,4}\corrauth{jiasm@ihep.ac.cn}
  \and Z. S. Yuan\inst{1,2,3}\email{zsyuan@nao.cas.cn}
  \and M. T. Shen\inst{5}\email{shenmt@stu.xmu.edu.cn}
  \and Y. Chen\inst{4}\email{ychen@ihep.ac.cn}
  \and C. K. Li\inst{4}\email{lick@ihep.ac.cn}
  \and C. Ge\inst{5}\email{chongge@xmu.edu.cn}
  \and L. M. Song\inst{3,4}\email{songlm@ihep.ac.cn}
  \and H. Feng\inst{4}\email{hfeng@ihep.ac.cn}
  \and J. Guan\inst{4}\email{jguan@ihep.ac.cn}
  \and C. C. Jin\inst{1}\email{ccjin@nao.cas.cn}
  \and C. Z. Liu\inst{4}\email{liucz@ihep.ac.cn}
  \and Y. Liu\inst{1}\email{liuyuan@nao.cas.cn}
  \and S. N. Zhang\inst{4}\email{zhangsn@ihep.ac.cn}
  \and H. S. Zhao\inst{4}\email{zhaohs@ihep.ac.cn}
}
\authorrunning{Wen et al.}

\institute{
National Astronomical Observatories, Chinese Academy of Sciences, A20 Datun Road, Chaoyang District, Beijing 100101, China
\and CAS Key Laboratory of FAST, NAOC, Chinese Academy of Sciences
\and University of Chinese Academy of Sciences, Beijing 100049, China
\and State Key Laboratory of Particle Astrophysics, Institute of High Energy Physics, Chinese Academy of Sciences, Beijing 100049, China.
\and Department of Astronomy, Xiamen University, Xiamen, Fujian 361005, China
}

\date{Received , accepted}


\abstract{ 

The evolutionary processes of galaxy clusters influence the properties
of their member galaxies. We present a joint X-ray--optical analysis
of the galaxy cluster WHY J050106.2$+$013714 at $z_{\rm
  c}=0.151$. X-ray observations with the \textit{Einstein} Probe
Follow-up X-ray Telescope indicate that the cluster is dynamically
young. The cluster displays an average X-ray temperature of
$2.8^{+0.4}_{-0.3}$ keV and a total luminosity of
9.4$\pm0.3\times10^{43}$ erg s$^{-1}$, consistent with the scaling
relation of typical disturbed clusters. Remarkably, the cluster hosts
a multi-merging brightest cluster galaxy (BCG) system composed of six
massive galaxies, with a total stellar mass of
$1.16\times10^{12}M_{\odot}$. We detected a well-defined intracluster
light component extending to a size of 310 kpc. A systematic search
for merging BCGs in the DESI Legacy Surveys reveals that this
sextuple-merging BCG is extremely rare in the local
Universe. Additionally, other merging BCGs are also likely to form in
moderately disturbed clusters, which provides valuable insights into
the formation of BCGs.}

\keywords{galaxies: clusters: individual --- X-rays: galaxies:
  clusters --- galaxies: elliptical and lenticular, cD}

\maketitle
\nolinenumbers
\section{Introduction}

As the largest gravitationally bound systems in the Universe, clusters
of galaxies reside at the density peaks of the large-scale structure
and serve as excellent tracers for constraining cosmological
parameters \citep[see the review by][]{aem11}. Galaxy clusters consist
of hot intracluster medium (ICM), tens to thousands of galaxies, and a
dominated dark matter component. Their enormous mass and high density
make them ideal laboratories for investigating the evolution of
galaxies \citep{dre80,wtc12,swx+24} and the physical mechanism of ICM
\citep{sar86,fgg+12}.

In the paradigm of structure formation, galaxy clusters assemble
through hierarchical process \citep{pee93}, i.e., continuous
merging of nearby structures. This formation process
implies that clusters have various evolutionary stages, each
exhibiting different observational features. Dynamically relaxed
clusters usually exhibit a bright X-ray-cool core and a regular
morphology \citep{vkf+06,jbp+08}, while dynamically young (disturbed)
clusters show a low X-ray surface brightness and a disturbed
morphology \citep{crb+07,hmr+10,ajf+17}.
The dynamical state of clusters plays a critical role in the evolution
of the member galaxies. For example, satellite galaxies in more
relaxed clusters usually have a lower star formation rate, an older
stellar population, and less disturbed morphologies
\citep{rp17,sr19,vcn+25}. There are more bright satellite galaxies in
disturbed clusters than in relaxed ones
\citep{wh13,mrd+20,arm+25}. The merging process between clusters can
enhance the ram-pressure stripping effect on the satellite galaxies,
resulting in the formation of more jellyfish galaxies
\citep{ocn+12,mer+16,ek19}.

In particular, central galaxies, referred to as brightest cluster
galaxies (BCGs), exhibit the strongest connections to the evolution of
their host clusters, including the correlations with cluster total
mass and dynamical state, the alignment between BCG and cluster
orientations, and the formation of intracluster light (ICL;
\citealt{whl12,cdv+14,yw22}).
BCGs have higher luminosities \citep{wh13,zkk+25}, larger sizes
\citep{ccl+13}, and greater radio powers \citep{yhw16} in more relaxed
clusters. Star-forming BCGs are preferentially found in relaxed
clusters, where X-ray-cool cores provide a steady supply of cold gas
\citep{obp+08,ree+12}.

\begin{figure*}
\sidecaption
\includegraphics[width=12cm]{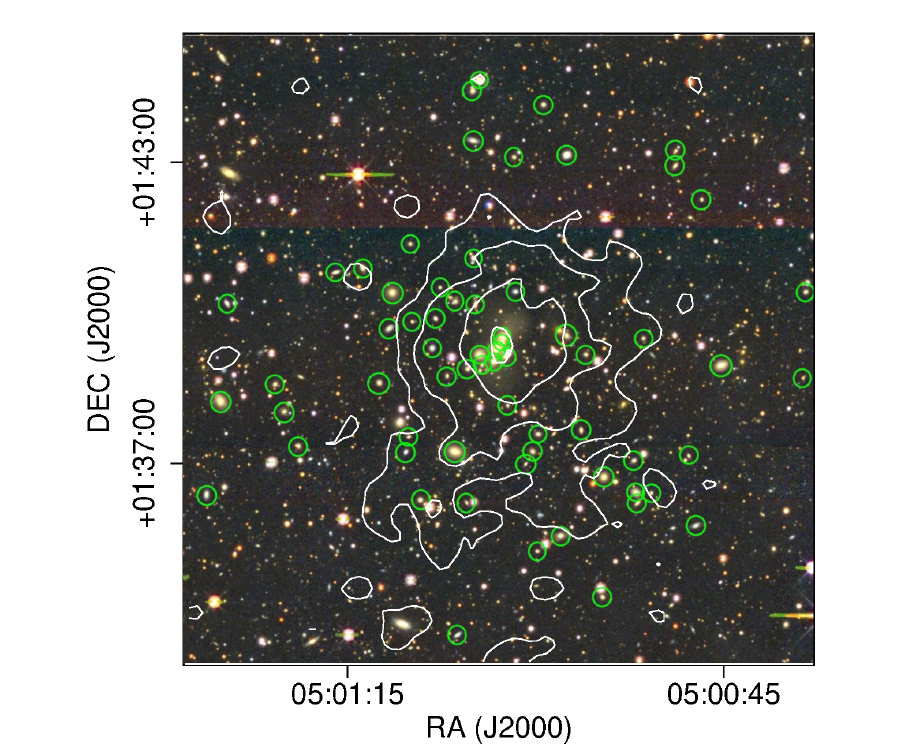}
\caption{Color image of the cluster WHY J0501$+$01 from DESI Legacy
  Surveys.  Open circles mark the member galaxy candidates selected by
  photometric redshift.  The size of the image is 2 Mpc$\times$2
  Mpc. The contour is X-ray emission from EP-FXT. }
\label{opticalimg}
\end{figure*}

The unique properties of BCGs provide observational support for the
hypothesis that merging processes drive BCG growth
\citep{bhs+07}. During the relaxation of a cluster, satellite galaxies
are inclined to migrate toward the cluster center due to dynamical
friction and are subsequently accreted by the BCG. This process leads
to the formation of a merging BCG system, ultimately resulting in its
growth.
However, whether major (a small mass ratio) or minor (a large mass
ratio) mergers are the dominant driver of BCG growth remains
controversial.
Some studies have found that BCG growth is dominated by frequent minor
mergers \citep{db07,ber09,yso+24}, while others argue that minor
mergers alone are insufficient and major mergers also play a
significant role \citep{gsg+17,kb23,mrp+23}.
Given the short timescales of major mergers, only a small fraction
($<5\%$) of BCGs are observed to be undergoing active major mergers,
predominantly in binary systems
\citep{mgh+08,lmd+09,bbh+21}. Multi-merging BCG systems are even
rarer; for instance, \citet{ggl+22} report a rare multi-merging BCG
system composed of seven galaxies with stellar masses ($M_{\star}$)
$>10^{10}~M_\odot$ and other two less massive galaxies. Currently, the
connection between the formation of merging BCGs and the evolutionary
stage of their host clusters has not been addressed.

Using the galaxy clusters identified from Dark Energy Spectroscopic
Instrument (DESI) Legacy Surveys \citep{wh24}, we obtained a sample of
ongoing merging BCG candidates in the local Universe. Among them, the
BCG system in the cluster WHY J050106.2$+$013714 (hereafter WHY
J0501$+$01) is particularly remarkable due to its major merger between
multiple massive galaxies. To investigate the properties of this
cluster, we conducted follow-up X-ray observations using the Follow-up
X-ray Telescope (FXT) on board \textit{Einstein} Probe (EP). In this
paper, we present a joint X-ray--optical analysis of WHY
J0501$+$01. Section 2 describes the observations and the data
reduction. Section 3 provides the X-ray results, and Sect. 4 focuses
on the multi-merging BCG system. Conclusions are presented in Sect. 5.

Throughout this paper, we assume a flat $\Lambda$ cold dark matter
cosmology with $H_0=70$ km~s$^{-1}$ Mpc$^{-1}$, $\Omega_m=0.3$, and
$\Omega_{\Lambda}=0.7$. 
\section{Observations}

\subsection{Optical data}

The cluster, WHY J0501$+$01, was initially identified from the
combined data of the Two Micron All Sky Survey, Wide-field Infrared
Survey Explorer, and SuperCOSMOS \citep{why18}.  \citet{wh24} later
re-identified it with improved measurements from the deeper data of
DESI Legacy Surveys \citep{dsl+19}. This cluster has a photometric
redshift of $z_{\rm c}=0.151\pm0.010$, an estimated radius of
$r_{500}=1.03$ Mpc (or $6.5'$) and a mass of $\log
M_{500}/M_{\odot}=14.50\pm0.20$. Here, $r_{500}$ denotes the radius
within which the mean density of the cluster is 500 times the critical
density of the Universe at the cluster redshift, while $M_{500}$
represents the total mass of the cluster enclosed within $r_{500}$.
Within a photometric redshift slice of $z_{\rm c}\pm0.04(1+z_{\rm
  c})$, we identified 64 member galaxy candidates with a stellar mass
of $\log M_\star/M_\odot\ge10$ within $r_{500}$, as illustrated in
Fig.~\ref{opticalimg} for their projected spatial distribution.  The
stellar masses of these galaxies were estimated from the
$z{\rm W1}$-band luminosities, with an uncertainty of 0.15 dex. Notably,
this nearby cluster lies in the coverage of the extended ROentgen
Survey with an Imaging Telescope Array (eROSITA)
All-Sky Survey but is not included in the eROSITA
X-ray cluster catalog \citep{blk+24}. Its optically estimated mass
exceeds that of 80\% eROSITA clusters at $z\sim0.15$. Thus, the
cluster is expected to have a relatively high total X-ray luminosity,
but its central brightness is too low for it to be detected by the
eROSITA. Here, our optical analysis uses the $griz$-band data from the
DESI Legacy Surveys with a 5\,$\sigma$ depth of $i\sim23.5$.

\subsection{X-ray observations and data reduction}

EP-FXT is one of the payloads on board the EP X-ray telescope, which
is a China-Europe collaboration mission lunched in January of 2024
\citep{2022yuanwm, 2025yuanwm}. It has a field of view of $\sim$
1$^{\circ}$ $\times$ 1$^{\circ}$, an angular resolution of $22''$
(half-power diameter), an effective area of approximately 600 cm$^2$
at 1 keV with two modules, and covers an energy range of 0.3–10
keV. Given the large field of view and low particle background
\citep{2025zheng,2025zhang}, EP-FXT is well suited for the study of
low-brightness diffuse sources.
WHY J0501$+$01 was detected by the EP-FXT during survey observations
in both identical modules of FXT-A and FXT-B. Later, we applied for a
Target of Opportunity (ToO) observation. All data used in this work
are in full-frame mode with the thin filter. The total good time
interval (GTI) is 11.7 ks, as listed in Table~\ref{obsdata}.

\begin{table}
\caption{EP-FXT observations of WHY J0501$+$01.}
\label{obsdata}
\centering
\begin{tabular}{cccc}
\hline\hline
ObsID & Date & GTI (ks)   & Obs. Type\\
\hline 
 11900008340 & 2024-09-30 & 1.5 & survey \\
 11900009170 & 2024-10-05 & 1.3 & survey \\
 11900053888 & 2025-01-19 & 8.9 & ToO \\
\hline
\end{tabular}
\end{table}

The data reduction followed the standard procedures of the FXT data
analysis software (FXTDAS version
1.20)$\footnote{https://epfxt.ihep.ac.cn/analysis}$ \citep{2025zhao,
  2025li}. We used the pipeline tool $fxtchain$ to generate the
calibrated and cleaned event files, $xselect$ to extract the images
and spectra, and $fxtarfgen$ and $fxtrmfgen$ to generate the response
files of arf and rmf, respectively. For more details, please refer to
\citet{2025zheng}.

\begin{figure}
\centering
\includegraphics[width = 0.48\textwidth]{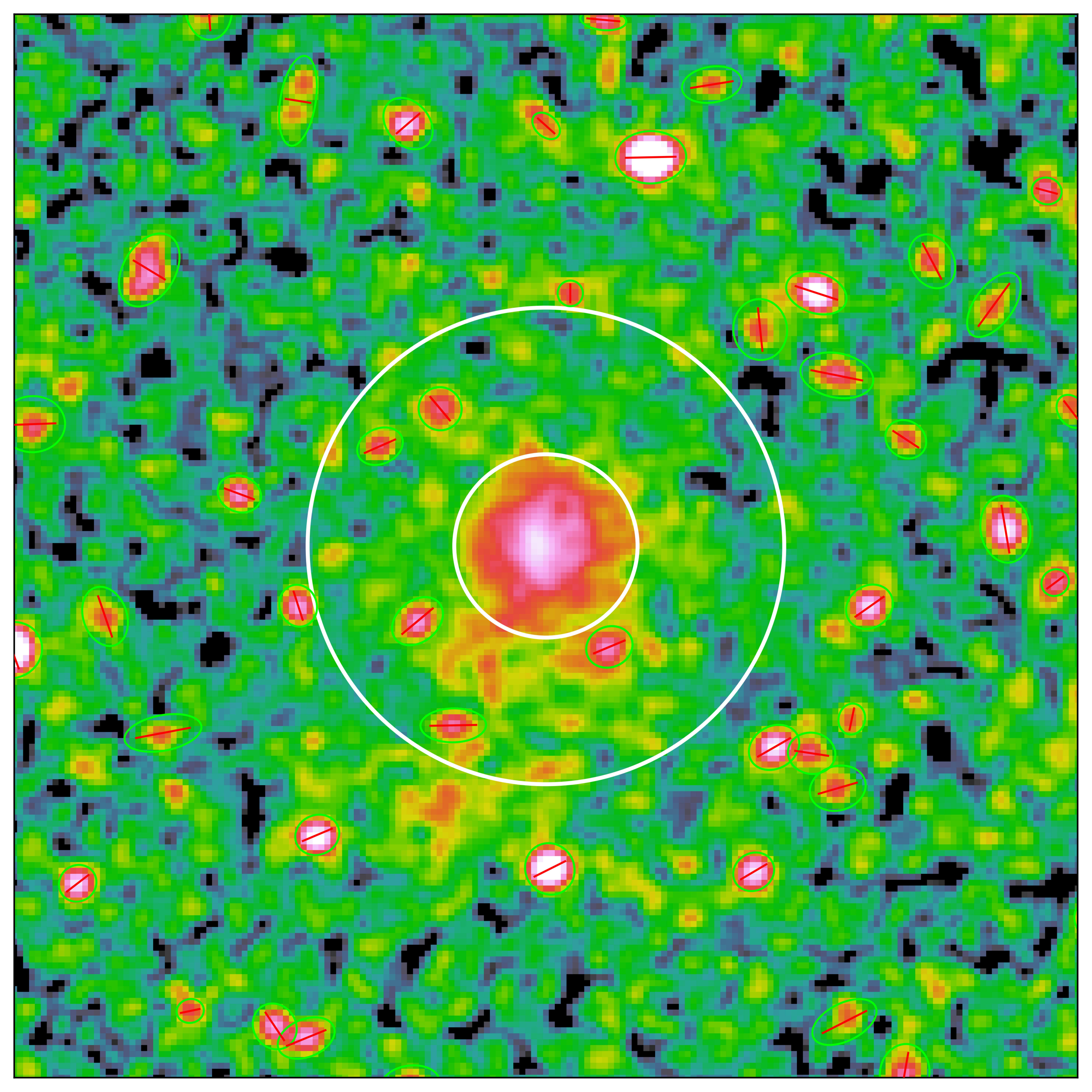}
\caption{Combined FXT X-ray image of WHY J0501$+$01 in the 0.5--10 keV
  band. The larger circle outlines the cluster region with a radius of
  $r_{500}$ ($6.5'$), and the smaller circle marks the inner region
  with a radius of $2.5'$.}
\label{Xrayimage}
\end{figure}

We combined the images of FXT-A and FXT-B for all these three
observations. The X-ray image of WHY J0501$+$01 is shown in
Fig.~\ref{Xrayimage}. After excluding the point sources, we plotted
the contour of the X-ray emission, which is overlaid on the optical
image in Fig.~\ref{opticalimg}. It exhibits substructures and an
elongated morphology, consistent with the distribution of massive
galaxies.

\begin{figure}
\centering
\includegraphics[width = 0.45\textwidth]{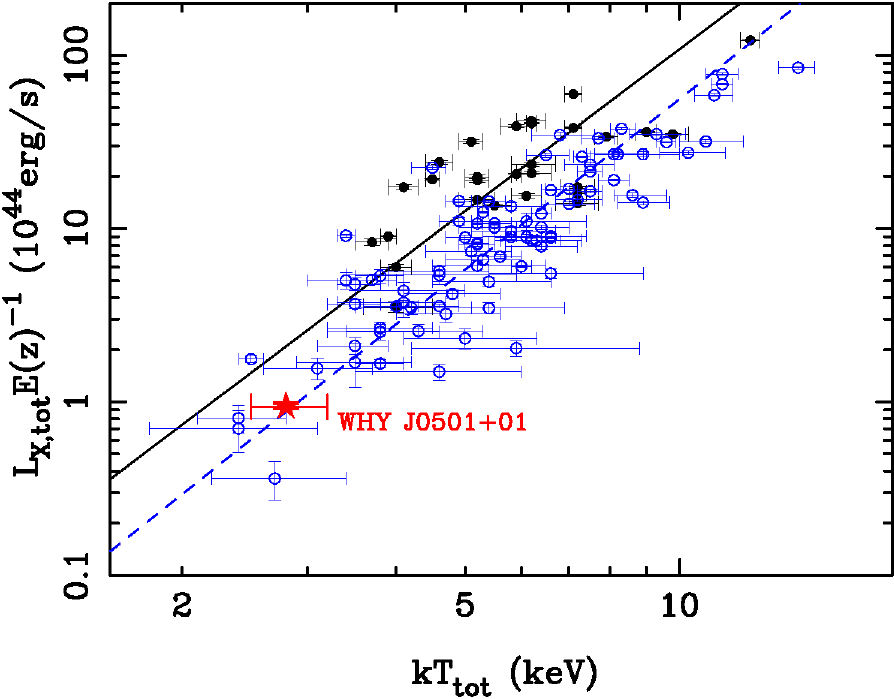}
\caption{X-ray luminosity and temperature of WHY J0501$+$01 (red star)
  in the $L_{\rm X}$--$kT$ diagram. Filled and open circles represent
  relaxed and disturbed clusters, respectively, from
  \citet{mgr+12}. The solid and dashed lines are scaling relations for
  relaxed and disturbed clusters, respectively.}
\label{LTrelation}
\end{figure}

\begin{figure*}
\centering
\includegraphics[width = 0.35\textwidth]{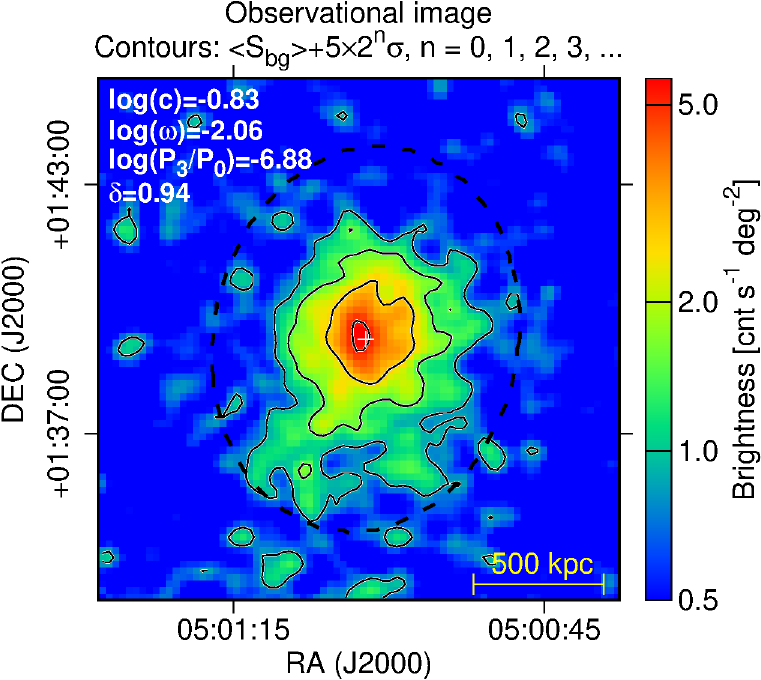}
\hspace{1mm}
\includegraphics[width = 0.313\textwidth]{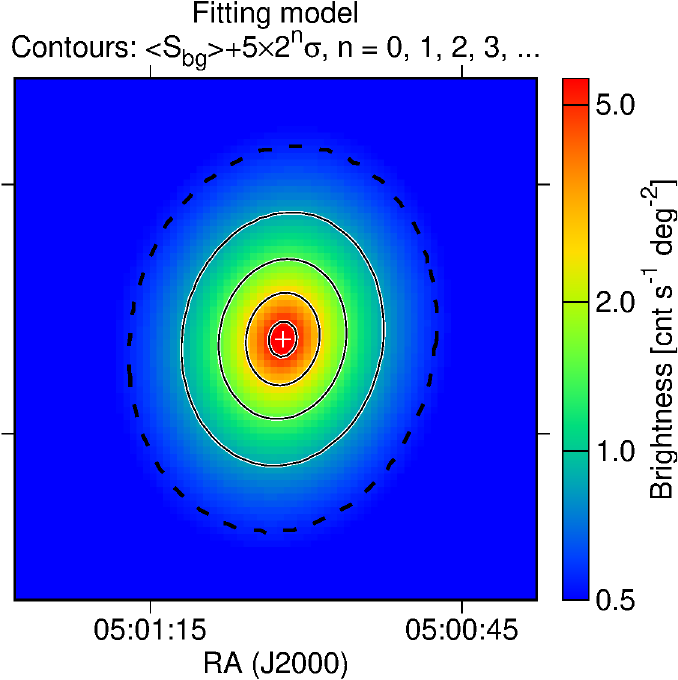}
\hspace{1mm}
\includegraphics[width = 0.3\textwidth]{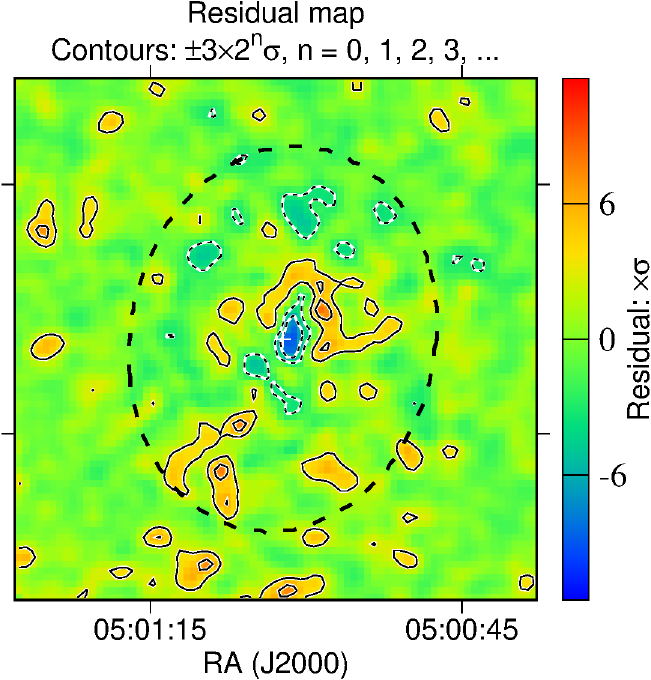}
\caption{\textit{Left}: X-ray image of WHY J0501$+$01 observed by the
  EP-FXT, with brightness contours. The white cross at the center
  denotes the fitted center, while the large dashed ellipse
  corresponds to the image region where we calculated the $\delta$
  parameter. The values of the four dynamical parameters are displayed
  in the upper-left corner. \textit{Middle}: Best-fitted $\beta$-model
  of WHY J0501$+$01. \textit{Right}: Residual map (observed minus
  model) of the cluster. Solid (dashed) contours denote positive
  (negative) value regions.}
\label{Dynimage}
\end{figure*}

\begin{figure*}
\sidecaption
\includegraphics[width=12cm]{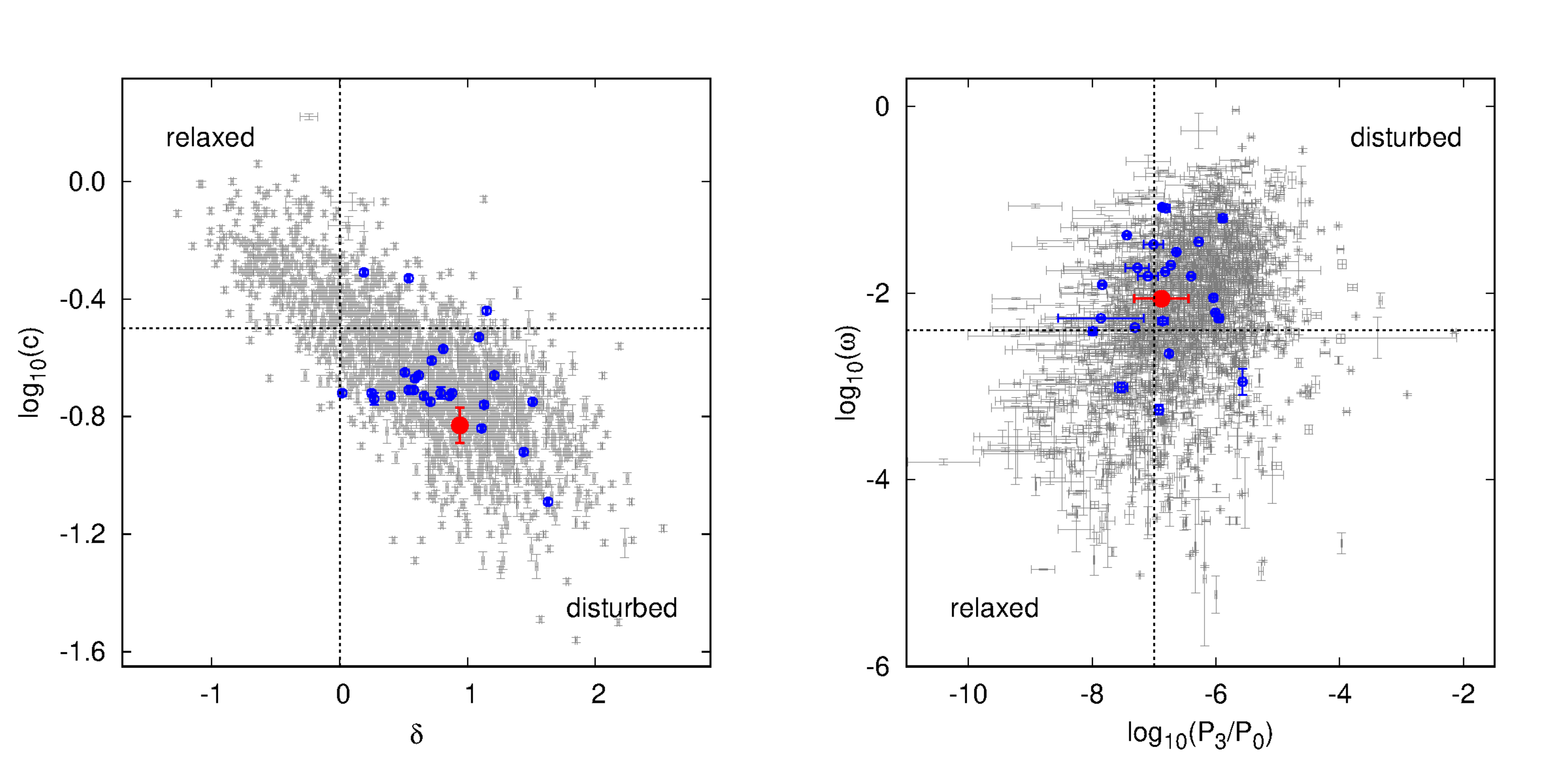}
\caption{Position of WHY J0501$+$01 (red dot) in the dynamical
  parameter space. The small gray dots are 1844 galaxy clusters from
  \citet{yhw22}, and the blue circles indicate other clusters hosting
  BCG mergers. The dotted lines denote the thresholds separating
  relaxed and disturbed clusters.}
\label{Dyndistr}
\end{figure*}

\section{X-ray results}

\subsection{Luminosity and temperature}

To study the X-ray properties of WHY J0501$+$01, we divided the
cluster into inner and outer regions, as marked by two white circles
in Fig.~\ref{Xrayimage}. The outer circle corresponds to the radius of
$r_{500}$ ($6.5'$), and the inner circle has a scale radius ($2.5'$)
assuming the Navarro--Frenk--White profile with a
concentration parameter from \citet{mmp+15}. A nearby clean region was
selected for background estimation. Spectral fitting for each region
was performed using XSPEC (version 12.14) with a model consisting of
(TBABS$\times$APEC). The cluster redshift ($z_{\rm c}=0.151$) and the
Galactic hydrogen column density \citep[$n_{\rm
    H}=7.68\times10^{20}{\rm cm}^{-2}$,][]{2013Willingale} were fixed,
while the temperature, metal abundance, and normalization were left as
free parameters. The fitting results are listed in
Table~\ref{fitresults}.
The spectral fitting gives an average temperature of
$2.8^{+0.4}_{-0.3}$ keV within the estimated $r_{500}$ ($6.5'$). In
the inner region ($<2.5'$), the temperature is $5.5\pm1.0$ keV, while
in the annular region between $2.5'$ and $6.5'$, it is
$2.0^{+0.4}_{-0.3}$ keV. The integration of X-ray flux derives a total
luminosity of $9.4\pm0.3\times10^{43}$ erg/s within $r_{500}$.

\begin{table*}
\footnotesize
\caption{Fitting results for WHY J0501$+$01 in the 0.5--10 keV band. }
\label{fitresults}
\centering
\begin{tabular*}{0.6\textwidth}{cccccc}
\hline\hline
Region & T & Abundance & Norm. & $\chi^2$/dof & Luminosity \\
 & [keV] & [Z$_{\odot}$] & [10$^{-3}$] & & [10$^{43}$ erg/s]\\
\hline
 0$-$2.5'  & $5.5\pm1.0$ & $<0.18$ & $1.00_{-0.05}^{+0.03}$ & 138.1/109 & $5.9\pm0.2$\\
 2.5$-$6.5'& $2.0_{-0.3}^{+0.4}$ & $<0.11$ & $1.25_{-0.15}^{+0.14}$ & 195.6/140 & $3.9\pm0.2$\\
 0$-$6.5'  & $2.8_{-0.3}^{+0.4}$ & $<0.07$ & $2.36_{-0.14}^{+0.08}$ & 340.1/219 & $9.4\pm0.3$\\
\hline
\end{tabular*}
\tablefoot{Uncertainties are 68\% confidence levels.}
\end{table*}

In Fig.~\ref{LTrelation} we compare the derived X-ray observables with
those for a sample of \textit{Chandra} clusters \citep{mgr+12}, whose
dynamical states are classified as relaxed or disturbed. We find that
the X-ray luminosity and temperature are consistent with the $L_{\rm
  X}$--$kT$ scaling relation of disturbed clusters, indicating that
WHY J0501$+$01 is a possible disturbed cluster.

\subsection{Dynamical state}

We further assessed the dynamical properties of WHY J0501$+$01 from
the structures of X-ray emission. Previously, \citet{yh20} calculated
four kinds of dynamical parameters --- namely concentration index
($c$), centroid shift ($\omega$), power ratio $P_3/P_0$, and
morphology index ($\delta$) --- for 964 galaxy clusters with
\textit{Chandra} image data. Later, they expanded the sample to 1844
clusters with \textit{XMM-Newton} data \citep{yhw22}. The
concentration index $c$ \citep[e.g.,][]{srt+08} quantifies the
relative central density of a cluster: relaxed clusters typically host
high-density cool cores, resulting in a high central concentration,
whereas merging clusters often have a lower central concentration.
The centroid shift parameter $\omega$ \citep{mfg93,pfb+06} measures
the centroid variations in different aperture sizes and is defined as
the standard deviation of the different center shifts. In relaxed
clusters, the cool core and centroid are generally well aligned,
whereas merging clusters often exhibit significant offsets between
their brightness peaks and centroids.
The power ratio $P_3/P_0$ \citep[e.g.,][]{bt95} measures the degree of
asymmetry or substructure in the X-ray surface brightness distribution
of a cluster. Relaxed clusters exhibit smooth, symmetric profiles with
minimal substructures, while merging clusters show significant
deviations due to the merger process.
Considering that the three parameters discussed above are all
calculated within a fixed circular aperture (e.g., a radius of 500
kpc), \citet{yh20} introduced a new morphological index $\delta$ to
characterize cluster dynamical states by accounting for its actual
size and morphology (see the large dashed ellipse in
Fig.~\ref{Dynimage}).

The left panel of Fig.~\ref{Dynimage} shows the observed X-ray image
used for analyzing the dynamical properties of WHY J0501$+$01. A
prominent tail-like structure is visible in the southern part of the
image, suggesting that the cluster is undergoing a recent
merger. Following the methodology of \citet{yh20} and \citet{yhw22},
we smoothed the X-ray image using a Gaussian function with a full
width at half maximum corresponding to 30 kpc, and then computed the
four dynamical parameters ($c$, $\omega$, $P_3/P_0$, and $\delta$) of
WHY J0501$+$01, which are displayed in the upper left corner.

To reveal the substructures of the WHY J0501$+$01, we fitted its X-ray
surface brightness profile with a two-dimensional $\beta$-model
\citep[e.g.,][]{yh20}. The middle panel of Fig.~\ref{Dynimage} shows
the best-fitted $\beta$-model, and the right panel displays the
residual map obtained by subtracting the model from the observed
image. The residual map reveals two distinct features. One is the
spiral-like substructure (solid contour) in the central region of the
cluster, indicating that the cluster core is sloshing
\citep[e.g.,][]{mg08}. The other is the southern substructure
corresponding to the gas tail visible in the observed image. To
quantify the amount of substructures, we measured the absolute
deviation of the photon counts in the residual image within the dashed
ellipse, and divided it by the total photon counts in the model
image. This yields a ratio of 0.17.

Figure~\ref{Dyndistr} shows the position of WHY J0501$+$01 in the
dynamical parameter space, together with the 1844 galaxy clusters from
\citet{yhw22}. Using the distribution of a complete sample of 125
galaxy clusters \citep[see Fig. 7 in][]{yh20}, we defined the
following thresholds to distinguish relaxed and disturbed clusters:
${\rm log}_{10}(c)=-0.5$, ${\rm log}_{10}(\omega)=-2.4$, ${\rm
  log}_{10}(P_3/P_0)=-7.0$, and $\delta=0$, as indicated by the dotted
lines in the figure.  Both panels confirm that WHY J0501$+$01 lies in
the disturbed regime.

\section{A remarkable BCG system}

\begin{figure*}
\centering
\includegraphics[width = 0.48\textwidth]{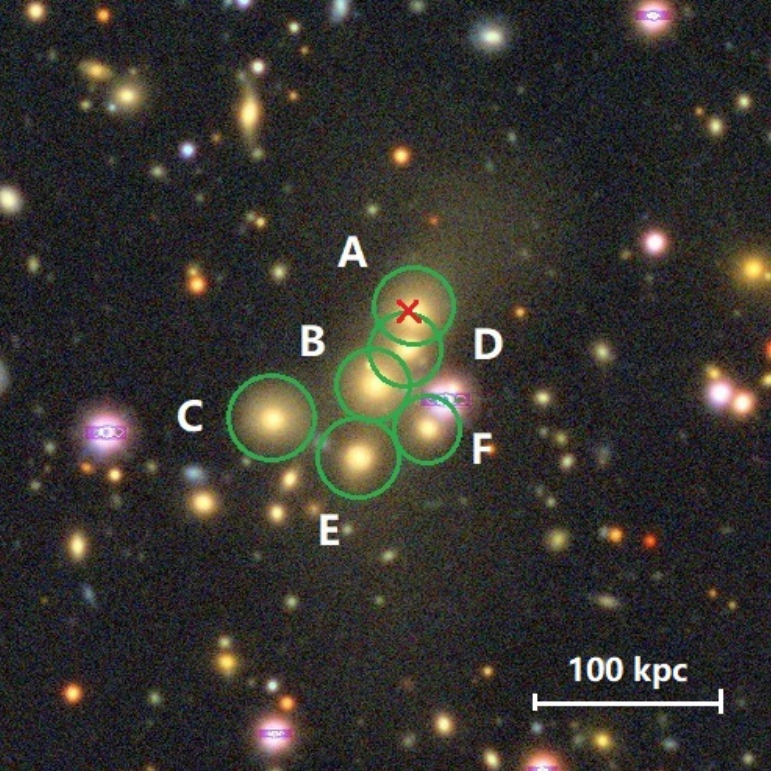}\hspace{1.mm}
\includegraphics[width = 0.48\textwidth]{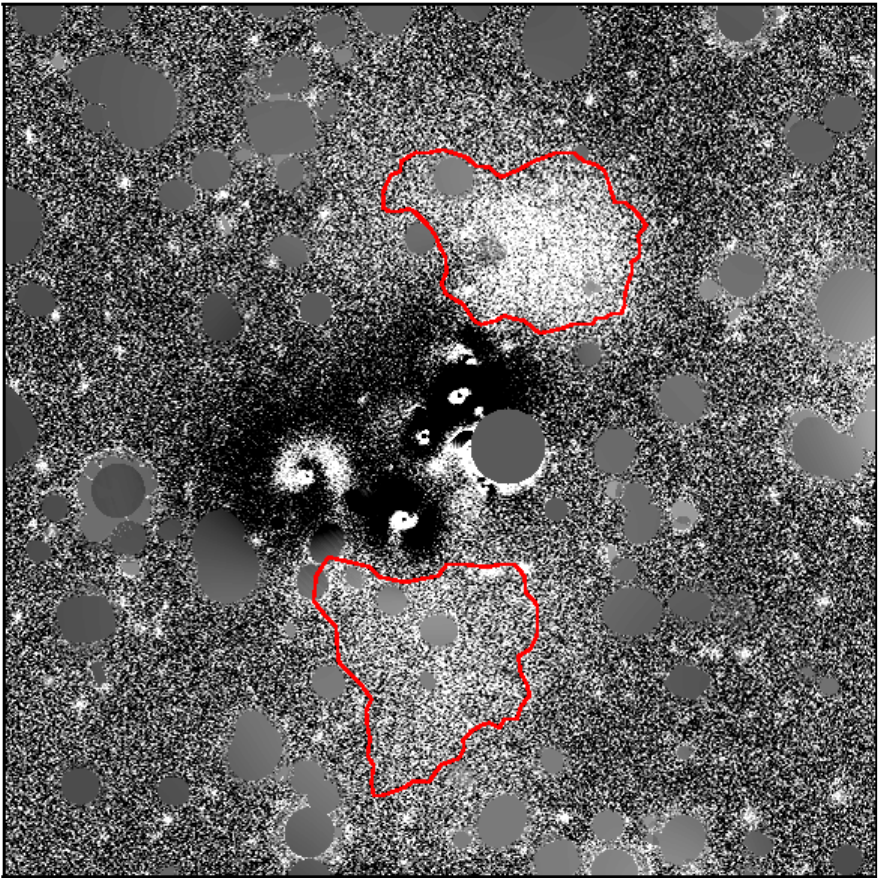}
\caption{\textit{Left}: Image of the BCG system in WHY J0501$+$01. The
  six galaxies involved in the BCG merger are labeled. The red cross
  marks the position of the X-ray emission peak. \textit{Right}: Same
  region after subtracting the model of the six labeled galaxies
  (other sources are masked). The solid lines represent the boundaries
  of ICL above 3\,$\sigma$.}
\label{BCG}
\end{figure*}

An intriguing system is observed in the core of WHY J0501$+$01 (left
panel of Fig.~\ref{BCG}), where six massive galaxies form a compact
group (Table~\ref{bcgdata}) and five of them have a stellar mass of
$\log M_\star/M_\odot\ge11$. If they merge, the newly formed BCG would
have a stellar mass of $1.16\times10^{12}~M_{\odot}$. Assuming the
cluster mass remains nearly constant during the merger, the final
stellar mass is higher than the $M_{\star,\rm BCG}$--$M_{500}$
relation \citep{blm+16} by 2.6\,$\sigma$.  Notably, the most massive
galaxy (labeled A) is offset from the system center but coincides with
the X-ray emission peak. Diffuse stellar components are visible in the
northern and southern directions, providing direct evidence that this
system is undergoing mergers. Clearer features can be shown in the
model-subtracted image (see Sect.~\ref{icl}). We designated this
system a sextuple-merging BCG.  Applying a similar calculation in
\citet{ggl+22}, we estimated the mean merging time $\langle T_{\rm
  merge}\rangle$ of the A, C, D, E, and F galaxies with the central B
galaxy, obtaining $\langle T_{\rm merge}\rangle$ in the range
$0.89\sim1.92$ Gyr.

\begin{table*}
\caption{Sources in the multi-merging BCG system of WHY J0501$+$01.}
\label{bcgdata}
\centering
\begin{tabular*}{0.95\textwidth}{ccccccccc}
\hline\hline
Name & R.A. & Dec. & $g$ & $r$ & $i$ & $z$ & $\log M_{\star}/M_\odot$ & D$_{\rm p}$ \\
(1) & (2) & (3) & (4) & (5) & (6) & (7) & (8) & (9)\\ 
\hline
 A & 75.26009 & 1.65825 & 17.479$\pm$0.005& 16.395$\pm$0.002& 15.994$\pm$0.001& 15.689$\pm$0.001& 11.43& 0.000\\
 B & 75.26217 & 1.65444 & 17.590$\pm$0.005& 16.489$\pm$0.005& 16.074$\pm$0.005& 15.765$\pm$0.005& 11.42& 0.041\\
 C & 75.26785 & 1.65249 & 17.479$\pm$0.005& 16.421$\pm$0.005& 16.024$\pm$0.005& 15.734$\pm$0.005& 11.38& 0.092\\
 D & 75.26044 & 1.65639 & 17.955$\pm$0.005& 16.906$\pm$0.005& 16.476$\pm$0.005& 16.165$\pm$0.005& 11.21& 0.018\\
 E & 75.26312 & 1.65040 & 17.949$\pm$0.005& 16.858$\pm$0.005& 16.451$\pm$0.005& 16.143$\pm$0.005& 11.20& 0.080\\
 F & 75.25932 & 1.65201 & 18.808$\pm$0.006& 17.755$\pm$0.003& 17.331$\pm$0.002& 17.033$\pm$0.002& 10.81& 0.059\\
ICL&   --     &  --     & 18.622$\pm$0.005& 17.859$\pm$0.005& 17.516$\pm$0.005& 17.145$\pm$0.005& 11.76& --   \\
\hline
\end{tabular*}
\tablefoot{
The columns are (1) source name;
(2)--(3) R.A. (J2000) and Dec. (J2000) in degrees;
(4)--(7) magnitudes in $griz$ bands, respectively;
(8) logarithm of stellar mass;
(9) projected distance to the most massive galaxy (in Mpc).
}
\end{table*}

\subsection{How rare are multi-merging BCGs and their dynamical environments?}

To assess the rareness of the sextuple-merging BCG, we systematically
searched for BCG mergers in the DESI Legacy surveys using clusters and
their massive member galaxy candidates with a stellar mass of $\log
M_\star/M_\odot\ge11$ \citep{wh24}. To resolve closely interacting
systems, we restricted our analysis to 52,803 nearby clusters at
$z_{\rm c}\le0.2$. Cross-matching with other X-ray or optical cluster
catalogs indicates that this sample is about 90\% complete for
clusters with $\log M_{500}/M_{\odot}\ge14$. Two close galaxies were
regarded as a potential binary merger if their separation is
sufficiently small. A multiple merger is defined as a compact group of
several galaxies, where each galaxy has a small separation from at
least one other member.  We applied a friend-of-friend algorithm to
these massive member galaxy candidates, starting from the central
BCGs. The projected linking length was set to 50 kpc
\citep{whl09,gsg+17}, and galaxies linked by this criterion are
considered as members of BCG mergers. Finally, we identified 2233 BCGs
in binary mergers, 70 in triple mergers, 12 in quadruple mergers, and
only one in a quintuple merger (i.e., WHY J0501$+$01, in which five of
six member galaxies have $\log M_\star/M_\odot\ge11$).  Note that not
all identified mergers exhibit distinct merging features, such as
asymmetric morphologies or diffuse tidal stellar structures, in the
optical images. Some may represent early-stage mergers or projected
alignments rather than true interactions \citep{mgh+08,lmd+09}. Thus,
these numbers represent upper limits on the total count of BCGs
undergoing strong mergers. Considering the large complete cluster
sample, we concluded that the quintuple merger in WHY J0501$+$01 thus
represents an exceptionally rare system.

The formation of BCG mergers relates to the evolution of their host
clusters. We have demonstrated that the sextuple-merging BCG resides
in a dynamically disturbed cluster. To investigate the dynamical
environment of other merging BCGs in the DESI Legacy Surveys, we
cross-matched the 2316 potential merging BCGs with 1844 X-ray clusters
observed by \textit{Chandra} and \textit{XMM-Newton}
\citep{yh20,yhw22}. This yielded 26 clusters with dynamical
parameters, as shown in Fig.~\ref{Dyndistr}.
The clusters hosting these merging BCGs exhibit moderate dynamical
parameters, with approximately 80\% of them in the range
$-0.9<\log(c)<-0.5$, $0.2<\delta<1.2$, $-3.0<\log(\omega)<-1.3$ and
$-7.8<\log(P_3/P_0)<-5.9$. Among these parameters, the concentration
index shows a narrower distribution with a peak at $\log(c)\sim-0.70$,
compared to the broader distributions of $\delta$, $\omega$, and
$P_3/P_0$.
If taking the empirical thresholds to distinguish relaxed and
disturbed clusters, we found that the merging BCGs predominantly
reside in disturbed clusters. This observed distribution can be
explained by the coevolution of BCGs and their host clusters. When two
clusters undergo violent merging, the newly formed cluster exhibits an
extremely disturbed morphology, and the massive BCGs within it possess
a large relative velocity. Under such conditions, their merger is
unlikely to occur.
As relaxation of the cluster, massive BCGs sink toward the cluster
center with a low relative velocity, enabling the formation of a BCG
merger. Once the cluster reaches a fully relaxed state, the merging
process between massive BCGs is generally complete.

\subsection{ICL}
\label{icl}

The image of the DESI Legacy Surveys clearly shows diffuse ICL
surrounding the merging BCG system in WHY J0501$+$01 (left panel of
Fig.~\ref{BCG}). To isolate and characterize the ICL component, we
performed two-dimensional structural modeling of the six massive
member galaxies and subtracted their best-fit models from the original
images. The GALFIT code \citep{phi+02} was applied to the $griz$-band
images and their corresponding point spread function images. Each
image was cropped to $600 \times 600$ pixels ($157.2 \times 157.2$
arcsec$^2$), ensuring full coverage of the target galaxies while
retaining sufficient surrounding area for reliable sky background
estimation. During the fitting process, we generated mask images using
Source Extractor Python (SEP) to remove contaminating sources outside
the target galaxies. The six member galaxies were modeled
simultaneously using six S\'ersic functions, with initial parameters
including central position, magnitude, effective radius ($R_{\rm e}$),
axis ratio ($b/a$), and position angle. These parameters were refined
iteratively through multiple fitting runs. We first fitted the
$z$-band image, which has the best data quality. For the $g$, $r$, and
$i$ bands, the structural parameters were initialized using the
best-fit values from the $z$ band and further optimized through
iterative GALFIT fitting, while the initial magnitudes were adopted
from the DESI Legacy Surveys (see Table~\ref{bcgdata}). The GALFIT
outputs include a best-fit model and a residual image (right panel of
Fig.~\ref{BCG}). In the residual image, the C galaxy exhibits a spiral
structure, while diffuse excess structures are visible near the cores
of the B, D and F galaxies.
The ICL is clearly detected to the north of the A galaxy and to the
south of the E galaxy. These features suggest that the six galaxies
are undergoing multiple mergers rather than being a chance
coincidence. Due to the relatively low brightness of ICL, only a
fraction of the pixels have values exceeding 3\,$\sigma$. To quantify
the spatial extent of ICL, we smoothed the image using a Gaussian
kernel with a standard deviation of 3 pixels. Subsequently, we
obtained the 3\,$\sigma$ contours (solid lines) for the northern and
southern ICL components. The maximum projected extent of the ICL,
measured from the south to the north, is 310 kpc.

The total ICL luminosity was estimated by integrating the residual
flux after subtracting the sky background. Due to poor modeling of
galaxy cores, the ICL brightness within 1\,$R_{\rm e}$ of each galaxy
was set to the average value measured within the red boundary
regions. For the masked regions, the ICL brightness was set to the
average value of the surrounding unmasked regions. We further
restricted the integration to a circular region with a radius of 200
pixels centered on the image.
From the GALFIT results, we derived the magnitudes of the ICL in each
band. The extinction-corrected $r$-band luminosity (adopting
$E(B-V)=0.098$, $R_r=2.165$; \citealt{Zhou_2025}) was then converted
to stellar mass using a mass-to-light ratio calibrated from the $r-z$
color, $M_\star/L$=10$^{[-0.241+0.780\,(r-z)]}$, adjusted to a
Chabrier initial mass function \citep{Bell_2003}.  The derived stellar
mass of the ICL is $5.7\times10^{11}M_\odot$, accounting for $\sim
18\%$ of the total stellar mass of the cluster. This ICL fraction is
consistent with previous measurements for disturbed clusters
\citep{jrh+25}.

\section{Conclusions}

Using observations from EP-FXT, we identified a new X-ray cluster with
an optical counterpart, WHY J0501$+$01, at $z_{\rm c}=0.151$. This
cluster exhibits an average X-ray temperature of $2.8^{+0.4}_{-0.3}$
keV and a total luminosity of $9.4\pm0.3\times10^{43}$ erg s$^{-1}$,
consistent with the established scaling relations for dynamically
disturbed galaxy clusters. Further dynamical analysis confirmed that
the cluster is unrelaxed, exhibiting a disturbed morphology.
Remarkably, the cluster hosts a compact group of six massive galaxies
with a total stellar mass of $1.16\times10^{12}~M_{\odot}$ that form a
sextuple-merging BCG system. After subtracting smooth models for these
galaxies, we clearly detected the ICL extending to a size of 310
kpc. A systematic search for merging BCGs in the DESI Legacy Surveys
suggests that such a multi-merging BCG system in WHY J0501$+$01 is
exceptionally rare in the local Universe and provides valuable
insights into the formation of BCGs and ICL. Moreover, other merging
BCGs are also likely found in clusters with unrelaxed dynamical
states, suggesting a correlation between the formation of BCG mergers
and the dynamical evolution of clusters.

\begin{acknowledgements}

We thank Prof. Han Jinlin and the referee for valuable comments that
helped to improve the paper.  The authors are supported by the
International Partnership Program of Chinese Academy of Sciences
(Grant No. 013GJHZ2024015FN), the National Astronomical Observatories
of the Chinese Academy of Sciences (No. E4ZR0506), the National Key
R\&D Program of China No. 2025YFF0511104 and the National Natural
Science Foundation of China (Grant No. 12373007, 12422302).  We also
acknowledge the support of the science research grants from the China
Manned Space Project with Numbers CMS-CSST-2025-A04. ZSY acknowledges
the support of the National SKA Program of China (Grant
No. 2022SKA0120103).

EP is a space mission supported by Strategic Priority Program 
on Space Science of Chinese Academy of Sciences, in collaborationwith 
ESA, MPE and CNES (Grant No. XDA15052100). 
The DESI Legacy Imaging Surveys consist of three individual and
complementary projects: the Dark Energy Camera Legacy Survey (DECaLS;
Proposal ID \#2014B-0404; PIs: David Schlegel and Arjun Dey), the
Beijing-Arizona Sky Survey (BASS; NOAO Prop. ID \#2015A-0801; PIs: Zhou
Xu and Xiaohui Fan), and the Mayall z-band Legacy Survey (MzLS;
Prop. ID \#2016A-0453; PI: Arjun Dey). DECaLS, BASS and MzLS together
include data obtained, respectively, at the Blanco telescope, Cerro
Tololo Inter-American Observatory, NSF’s NOIRLab; the Bok telescope,
Steward Observatory, University of Arizona; and the Mayall telescope,
Kitt Peak National Observatory, NOIRLab. Pipeline processing and
analyses of the data were supported by NOIRLab and the Lawrence
Berkeley National Laboratory (LBNL). The Legacy Surveys project is
honored to be permitted to conduct astronomical research on Iolkam
Du’ag (Kitt Peak), a mountain with particular significance to the
Tohono O’odham Nation.
\end{acknowledgements}

\bibliographystyle{aa}
\bibliography{cluster}

@ARTICLE{arm+25,
       author = {{Ahad}, Syeda Lammim and {Reid}, Rashaad and {Mpetha}, Charlie T. and {Taylor}, James E. and {Hildebrandt}, Hendrik and {Hudson}, Michael J. and {Chambers}, Kenneth C. and {de Boer}, Thomas and {Guerrini}, Sacha and {Guinot}, Axel and {Gwyn}, Stephen and {Kilbinger}, Martin and {Van Waerbeke}, Ludovic},
        title = "{Cluster properties as a function of dynamical state in the DESI Legacy x UNIONS surveys}",
      journal = {Submitted to ApJ},
         year = 2025,
        month = dec,
          eid = {arXiv:2512.14636},
        pages = {arXiv:2512.14636},
          doi = {10.48550/arXiv.2512.14636},
archivePrefix = {arXiv},
       eprint = {2512.14636},
 primaryClass = {astro-ph.GA},
       adsurl = {https://ui.adsabs.harvard.edu/abs/2025arXiv251214636A}
}

@ARTICLE{aem11,
       author = {{Allen}, Steven W. and {Evrard}, August E. and {Mantz}, Adam B.},
        title = "{Cosmological Parameters from Observations of Galaxy Clusters}",
      journal = {\araa},
         year = 2011,
        month = sep,
       volume = {49},
       number = {1},
        pages = {409-470},
          doi = {10.1146/annurev-astro-081710-102514},
archivePrefix = {arXiv},
       eprint = {1103.4829},
 primaryClass = {astro-ph.CO},
       adsurl = {https://ui.adsabs.harvard.edu/abs/2011ARA&A..49..409A}
}

@ARTICLE{ajf+17,
       author = {{Andrade-Santos}, Felipe and {Jones}, Christine and {Forman}, William R. and {Lovisari}, Lorenzo and {Vikhlinin}, Alexey and {van Weeren}, Reinout J. and {Murray}, Stephen S. and {Arnaud}, Monique and {Pratt}, Gabriel W. and {D{\'e}mocl{\`e}s}, Jessica and {Kraft}, Ralph and {Mazzotta}, Pasquale and {B{\"o}hringer}, Hans and {Chon}, Gayoung and {Giacintucci}, Simona and {Clarke}, Tracy E. and {Borgani}, Stefano and {David}, Larry and {Douspis}, Marian and {Pointecouteau}, Etienne and {Dahle}, H{\r{a}}kon and {Brown}, Shea and {Aghanim}, Nabila and {Rasia}, Elena},
        title = "{The Fraction of Cool-core Clusters in X-Ray versus SZ Samples Using Chandra Observations}",
      journal = {\apj},
         year = 2017,
        month = jul,
       volume = {843},
       number = {1},
          eid = {76},
        pages = {76},
          doi = {10.3847/1538-4357/aa7461},
archivePrefix = {arXiv},
       eprint = {1703.08690},
 primaryClass = {astro-ph.CO},
       adsurl = {https://ui.adsabs.harvard.edu/abs/2017ApJ...843...76A}
}

@ARTICLE{bbh+21,
       author = {{Banks}, K. and {Brough}, S. and {Holwerda}, B.~W. and {Hopkins}, A.~M. and {L{\'o}pez-S{\'a}nchez}, {\'A}. R. and {Phillipps}, S. and {Pimbblet}, K.~A. and {Robotham}, A.~S.~G.},
        title = "{Galaxy And Mass Assembly (GAMA): The Merging Potential of Brightest Group Galaxies}",
      journal = {\apj},
         year = 2021,
        month = nov,
       volume = {921},
       number = {1},
          eid = {47},
        pages = {47},
          doi = {10.3847/1538-4357/ac1c0a},
archivePrefix = {arXiv},
       eprint = {2108.05923},
 primaryClass = {astro-ph.GA},
       adsurl = {https://ui.adsabs.harvard.edu/abs/2021ApJ...921...47B}
}

@ARTICLE{blm+16,
       author = {{Bellstedt}, Sabine and {Lidman}, Chris and {Muzzin}, Adam and {Franx}, Marijn and {Guatelli}, Susanna and {Hill}, Allison R. and {Hoekstra}, Henk and {Kurinsky}, Noah and {Labbe}, Ivo and {Marchesini}, Danilo and {Marsan}, Z. Cemile and {Safavi-Naeini}, Mitra and {Sif{\'o}n}, Crist{\'o}bal and {Stefanon}, Mauro and {van de Sande}, Jesse and {van Dokkum}, Pieter and {Weigel}, Catherine},
        title = "{The evolution in the stellar mass of brightest cluster galaxies over the past 10 billion years}",
      journal = {\mnras},
         year = 2016,
        month = aug,
       volume = {460},
       number = {3},
        pages = {2862-2874},
          doi = {10.1093/mnras/stw1184},
archivePrefix = {arXiv},
       eprint = {1605.02736},
 primaryClass = {astro-ph.GA},
       adsurl = {https://ui.adsabs.harvard.edu/abs/2016MNRAS.460.2862B}
}

@ARTICLE{bhs+07,
       author = {{Bernardi}, Mariangela and {Hyde}, Joseph B. and {Sheth}, Ravi K. and {Miller}, Chris J. and {Nichol}, Robert C.},
        title = "{The Luminosities, Sizes, and Velocity Dispersions of Brightest Cluster Galaxies: Implications for Formation History}",
      journal = {\aj},
         year = 2007,
        month = apr,
       volume = {133},
       number = {4},
        pages = {1741-1755},
          doi = {10.1086/511783},
archivePrefix = {arXiv},
       eprint = {astro-ph/0607117},
 primaryClass = {astro-ph},
       adsurl = {https://ui.adsabs.harvard.edu/abs/2007AJ....133.1741B}
}

@ARTICLE{ber09,
       author = {{Bernardi}, Mariangela},
        title = "{Evolution in the structural properties of early-type brightest cluster galaxies at small lookback time and dependence on the environment}",
      journal = {\mnras},
         year = 2009,
        month = may,
       volume = {395},
       number = {3},
        pages = {1491-1506},
          doi = {10.1111/j.1365-2966.2009.14601.x},
archivePrefix = {arXiv},
       eprint = {0901.1318},
 primaryClass = {astro-ph.CO},
       adsurl = {https://ui.adsabs.harvard.edu/abs/2009MNRAS.395.1491B}
}

@ARTICLE{blk+24,
       author = {{Bulbul}, E. and {Liu}, A. and {Kluge}, M. and {Zhang}, X. and {Sanders}, J.~S. and {Bahar}, Y.~E. and {Ghirardini}, V. and {Artis}, E. and {Seppi}, R. and {Garrel}, C. and {Ramos-Ceja}, M.~E. and {Comparat}, J. and {Balzer}, F. and {B{\"o}ckmann}, K. and {Br{\"u}ggen}, M. and {Clerc}, N. and {Dennerl}, K. and {Dolag}, K. and {Freyberg}, M. and {Grandis}, S. and {Gruen}, D. and {Kleinebreil}, F. and {Krippendorf}, S. and {Lamer}, G. and {Merloni}, A. and {Migkas}, K. and {Nandra}, K. and {Pacaud}, F. and {Predehl}, P. and {Reiprich}, T.~H. and {Schrabback}, T. and {Veronica}, A. and {Weller}, J. and {Zelmer}, S.},
        title = "{The SRG/eROSITA All-Sky Survey. The first catalog of galaxy clusters and groups in the Western Galactic Hemisphere}",
      journal = {\aap},
         year = 2024,
        month = may,
       volume = {685},
          eid = {A106},
        pages = {A106},
          doi = {10.1051/0004-6361/202348264},
archivePrefix = {arXiv},
       eprint = {2402.08452},
 primaryClass = {astro-ph.CO},
       adsurl = {https://ui.adsabs.harvard.edu/abs/2024A&A...685A.106B}
}

@ARTICLE{bt95,
       author = {{Buote}, David A. and {Tsai}, John C.},
        title = "{Quantifying the Morphologies and Dynamical Evolution of Galaxy Clusters. I. The Method}",
      journal = {\apj},
         year = 1995,
        month = oct,
       volume = {452},
        pages = {522},
          doi = {10.1086/176326},
archivePrefix = {arXiv},
       eprint = {astro-ph/9502002},
 primaryClass = {astro-ph},
       adsurl = {https://ui.adsabs.harvard.edu/abs/1995ApJ...452..522B}
}

@ARTICLE{ccl+13,
       author = {{Carollo}, C. Marcella and {Cibinel}, Anna and {Lilly}, Simon J. and {Miniati}, Francesco and {Norberg}, Peder and {Silverman}, John D. and {van Gorkom}, Jacqueline and {Cameron}, Ewan and {Finoguenov}, Alexis and {Peng}, Yingjie and {Pipino}, Antonio and {Rudick}, Craig S.},
        title = "{The Zurich Environmental Study of Galaxies in Groups along the Cosmic Web. I. Which Environment Affects Galaxy Evolution?}",
      journal = {\apj},
         year = 2013,
        month = oct,
       volume = {776},
       number = {2},
          eid = {71},
        pages = {71},
          doi = {10.1088/0004-637X/776/2/71},
archivePrefix = {arXiv},
       eprint = {1206.5807},
 primaryClass = {astro-ph.CO},
       adsurl = {https://ui.adsabs.harvard.edu/abs/2013ApJ...776...71C}
}

@ARTICLE{crb+07,
       author = {{Chen}, Y. and {Reiprich}, T.~H. and {B{\"o}hringer}, H. and {Ikebe}, Y. and {Zhang}, Y.-Y.},
        title = "{Statistics of X-ray observables for the cooling-core and non-cooling core galaxy clusters}",
      journal = {\aap},
         year = 2007,
        month = may,
       volume = {466},
       number = {3},
        pages = {805-812},
          doi = {10.1051/0004-6361:20066471},
archivePrefix = {arXiv},
       eprint = {astro-ph/0702482},
 primaryClass = {astro-ph},
       adsurl = {https://ui.adsabs.harvard.edu/abs/2007A&A...466..805C}
}

@ARTICLE{cdv+14,
       author = {{Contini}, E. and {De Lucia}, G. and {Villalobos}, {\'A}. and {Borgani}, S.},
        title = "{On the formation and physical properties of the intracluster light in hierarchical galaxy formation models}",
      journal = {\mnras},
         year = 2014,
        month = feb,
       volume = {437},
       number = {4},
        pages = {3787-3802},
          doi = {10.1093/mnras/stt2174},
archivePrefix = {arXiv},
       eprint = {1311.2076},
 primaryClass = {astro-ph.CO},
       adsurl = {https://ui.adsabs.harvard.edu/abs/2014MNRAS.437.3787C}
}

@ARTICLE{dsl+19,
       author = {{Dey}, Arjun and {Schlegel}, David J. and {Lang}, Dustin and {Blum}, Robert and {Burleigh}, Kaylan and {Fan}, Xiaohui and {Findlay}, Joseph R. and {Finkbeiner}, Doug and {Herrera}, David and {Juneau}, St{\'e}phanie and {Landriau}, Martin and {Levi}, Michael and {McGreer}, Ian and {Meisner}, Aaron and {Myers}, Adam D. and {Moustakas}, John and {Nugent}, Peter and {Patej}, Anna and {Schlafly}, Edward F. and {Walker}, Alistair R. and {Valdes}, Francisco and {Weaver}, Benjamin A. and {Y{\`e}che}, Christophe and {Zou}, Hu and {Zhou}, Xu and {Abareshi}, Behzad and {Abbott}, T.~M.~C. and {Abolfathi}, Bela and {Aguilera}, C. and {Alam}, Shadab and {Allen}, Lori and {Alvarez}, A. and {Annis}, James and {Ansarinejad}, Behzad and {Aubert}, Marie and {Beechert}, Jacqueline and {Bell}, Eric F. and {BenZvi}, Segev Y. and {Beutler}, Florian and {Bielby}, Richard M. and {Bolton}, Adam S. and {Brice{\~n}o}, C{\'e}sar and {Buckley-Geer}, Elizabeth J. and {Butler}, Karen and {Calamida}, Annalisa and {Carlberg}, Raymond G. and {Carter}, Paul and {Casas}, Ricard and {Castander}, Francisco J. and {Choi}, Yumi and {Comparat}, Johan and {Cukanovaite}, Elena and {Delubac}, Timoth{\'e}e and {DeVries}, Kaitlin and {Dey}, Sharmila and {Dhungana}, Govinda and {Dickinson}, Mark and {Ding}, Zhejie and {Donaldson}, John B. and {Duan}, Yutong and {Duckworth}, Christopher J. and {Eftekharzadeh}, Sarah and {Eisenstein}, Daniel J. and {Etourneau}, Thomas and {Fagrelius}, Parker A. and {Farihi}, Jay and {Fitzpatrick}, Mike and {Font-Ribera}, Andreu and {Fulmer}, Leah and {G{\"a}nsicke}, Boris T. and {Gaztanaga}, Enrique and {George}, Koshy and {Gerdes}, David W. and {Gontcho}, Satya Gontcho A. and {Gorgoni}, Claudio and {Green}, Gregory and {Guy}, Julien and {Harmer}, Diane and {Hernandez}, M. and {Honscheid}, Klaus and {Huang}, Lijuan Wendy and {James}, David J. and {Jannuzi}, Buell T. and {Jiang}, Linhua and {Joyce}, Richard and {Karcher}, Armin and {Karkar}, Sonia and {Kehoe}, Robert and {Kneib}, Jean-Paul and {Kueter-Young}, Andrea and {Lan}, Ting-Wen and {Lauer}, Tod R. and {Le Guillou}, Laurent and {Le Van Suu}, Auguste and {Lee}, Jae Hyeon and {Lesser}, Michael and {Perreault Levasseur}, Laurence and {Li}, Ting S. and {Mann}, Justin L. and {Marshall}, Robert and {Mart{\'\i}nez-V{\'a}zquez}, C.~E. and {Martini}, Paul and {du Mas des Bourboux}, H{\'e}lion and {McManus}, Sean and {Meier}, Tobias Gabriel and {M{\'e}nard}, Brice and {Metcalfe}, Nigel and {Mu{\~n}oz-Guti{\'e}rrez}, Andrea and {Najita}, Joan and {Napier}, Kevin and {Narayan}, Gautham and {Newman}, Jeffrey A. and {Nie}, Jundan and {Nord}, Brian and {Norman}, Dara J. and {Olsen}, Knut A.~G. and {Paat}, Anthony and {Palanque-Delabrouille}, Nathalie and {Peng}, Xiyan and {Poppett}, Claire L. and {Poremba}, Megan R. and {Prakash}, Abhishek and {Rabinowitz}, David and {Raichoor}, Anand and {Rezaie}, Mehdi and {Robertson}, A.~N. and {Roe}, Natalie A. and {Ross}, Ashley J. and {Ross}, Nicholas P. and {Rudnick}, Gregory and {Safonova}, Sasha and {Saha}, Abhijit and {S{\'a}nchez}, F. Javier and {Savary}, Elodie and {Schweiker}, Heidi and {Scott}, Adam and {Seo}, Hee-Jong and {Shan}, Huanyuan and {Silva}, David R. and {Slepian}, Zachary and {Soto}, Christian and {Sprayberry}, David and {Staten}, Ryan and {Stillman}, Coley M. and {Stupak}, Robert J. and {Summers}, David L. and {Sien Tie}, Suk and {Tirado}, H. and {Vargas-Maga{\~n}a}, Mariana and {Vivas}, A. Katherina and {Wechsler}, Risa H. and {Williams}, Doug and {Yang}, Jinyi and {Yang}, Qian and {Yapici}, Tolga and {Zaritsky}, Dennis and {Zenteno}, A. and {Zhang}, Kai and {Zhang}, Tianmeng and {Zhou}, Rongpu and {Zhou}, Zhimin},
        title = "{Overview of the DESI Legacy Imaging Surveys}",
      journal = {\aj},
         year = 2019,
        month = may,
       volume = {157},
       number = {5},
          eid = {168},
        pages = {168},
          doi = {10.3847/1538-3881/ab089d},
archivePrefix = {arXiv},
       eprint = {1804.08657},
 primaryClass = {astro-ph.IM},
       adsurl = {https://ui.adsabs.harvard.edu/abs/2019AJ....157..168D}
}

@ARTICLE{db07,
       author = {{De Lucia}, Gabriella and {Blaizot}, J{\'e}r{\'e}my},
        title = "{The hierarchical formation of the brightest cluster galaxies}",
      journal = {\mnras},
         year = 2007,
        month = feb,
       volume = {375},
       number = {1},
        pages = {2-14},
          doi = {10.1111/j.1365-2966.2006.11287.x},
archivePrefix = {arXiv},
       eprint = {astro-ph/0606519},
 primaryClass = {astro-ph},
       adsurl = {https://ui.adsabs.harvard.edu/abs/2007MNRAS.375....2D}
}

@ARTICLE{dre80,
       author = {{Dressler}, A.},
        title = "{Galaxy morphology in rich clusters: implications for the formation and evolution of galaxies.}",
      journal = {\apj},
         year = 1980,
        month = mar,
       volume = {236},
        pages = {351-365},
          doi = {10.1086/157753},
       adsurl = {https://ui.adsabs.harvard.edu/abs/1980ApJ...236..351D}
}

@ARTICLE{ek19,
       author = {{Ebeling}, Harald and {Kalita}, Boris S.},
        title = "{Jellyfish: Ram Pressure Stripping As a Diagnostic Tool in Studies of Cluster Collisions}",
      journal = {\apj},
         year = 2019,
        month = sep,
       volume = {882},
       number = {2},
          eid = {127},
        pages = {127},
          doi = {10.3847/1538-4357/ab35d6},
archivePrefix = {arXiv},
       eprint = {1907.12781},
 primaryClass = {astro-ph.GA},
       adsurl = {https://ui.adsabs.harvard.edu/abs/2019ApJ...882..127E}
}

@ARTICLE{fgg+12,
       author = {{Feretti}, Luigina and {Giovannini}, Gabriele and {Govoni}, Federica and {Murgia}, Matteo},
        title = "{Clusters of galaxies: observational properties of the diffuse radio emission}",
      journal = {\aapr},
         year = 2012,
        month = may,
       volume = {20},
          eid = {54},
        pages = {54},
          doi = {10.1007/s00159-012-0054-z},
archivePrefix = {arXiv},
       eprint = {1205.1919},
 primaryClass = {astro-ph.CO},
       adsurl = {https://ui.adsabs.harvard.edu/abs/2012A&ARv..20...54F}
}

@ARTICLE{ggl+22,
       author = {{Geng}, Chao and {Ge}, Chong and {Lal}, Dharam V. and {Sun}, Ming and {Ji}, Li and {Xu}, Haiguang and {Liu}, Wenhao and {Hardcastle}, Martin and {Forman}, William and {Kraft}, Ralph and {Jones}, Christine},
        title = "{Chandra view of Abell 407: the central compact group of galaxies and the interaction between the radio AGN and the ICM}",
      journal = {\mnras},
         year = 2022,
        month = apr,
       volume = {511},
       number = {3},
        pages = {3994-4004},
          doi = {10.1093/mnras/stac355},
archivePrefix = {arXiv},
       eprint = {2202.03742},
 primaryClass = {astro-ph.GA},
       adsurl = {https://ui.adsabs.harvard.edu/abs/2022MNRAS.511.3994G}
}

@ARTICLE{gsg+17,
       author = {{Groenewald}, Dani{\`e}l N. and {Skelton}, Rosalind E. and {Gilbank}, David G. and {Loubser}, S. Ilani},
        title = "{The close pair fraction of BCGs since z = 0.5: major mergers dominate recent BCG stellar mass growth}",
      journal = {\mnras},
         year = 2017,
        month = jun,
       volume = {467},
       number = {4},
        pages = {4101-4117},
          doi = {10.1093/mnras/stx340},
archivePrefix = {arXiv},
       eprint = {1701.09012},
 primaryClass = {astro-ph.GA},
       adsurl = {https://ui.adsabs.harvard.edu/abs/2017MNRAS.467.4101G}
}

@ARTICLE{hmr+10,
       author = {{Hudson}, D.~S. and {Mittal}, R. and {Reiprich}, T.~H. and {Nulsen}, P.~E.~J. and {Andernach}, H. and {Sarazin}, C.~L.},
        title = "{What is a cool-core cluster? a detailed analysis of the cores of the X-ray flux-limited HIFLUGCS cluster sample}",
      journal = {\aap},
         year = 2010,
        month = apr,
       volume = {513},
          eid = {A37},
        pages = {A37},
          doi = {10.1051/0004-6361/200912377},
archivePrefix = {arXiv},
       eprint = {0911.0409},
 primaryClass = {astro-ph.CO},
       adsurl = {https://ui.adsabs.harvard.edu/abs/2010A&A...513A..37H}
}

@ARTICLE{jbp+08,
       author = {{Jia}, S.~M. and {B{\"o}hringer}, H. and {Pointecouteau}, E. and {Chen}, Y. and {Zhang}, Y.~Y.},
        title = "{XMM-Newton studies of a massive cluster of galaxies: RXC J2228.6+2036}",
      journal = {\aap},
         year = 2008,
        month = oct,
       volume = {489},
       number = {1},
        pages = {1-9},
          doi = {10.1051/0004-6361:200809699},
archivePrefix = {arXiv},
       eprint = {0806.1575},
 primaryClass = {astro-ph},
       adsurl = {https://ui.adsabs.harvard.edu/abs/2008A&A...489....1J}
}

@ARTICLE{jrh+25,
       author = {{Jim{\'e}nez-Teja}, Yolanda and {Rom{\'a}n}, Javier and {HyeongHan}, Kim and {V{\'\i}lchez}, Jose M. and {Dupke}, Renato A. and {Lopes}, Paulo Afr{\^a}nio Augusto and {Rich}, Robert Michael and {Caceres}, Osmin and {Li}, Chester},
        title = "{Deep view of the intracluster light in the Coma cluster of galaxies}",
      journal = {\aap},
         year = 2025,
        month = feb,
       volume = {694},
          eid = {A216},
        pages = {A216},
          doi = {10.1051/0004-6361/202452384},
archivePrefix = {arXiv},
       eprint = {2412.15328},
 primaryClass = {astro-ph.GA},
       adsurl = {https://ui.adsabs.harvard.edu/abs/2025A&A...694A.216J}
}

@ARTICLE{kb23,
       author = {{Kluge}, Matthias and {Bender}, Ralf},
        title = "{Minor Mergers Are Not Enough: The Importance of Major Mergers during Brightest Cluster Galaxy Assembly}",
      journal = {\apjs},
         year = 2023,
        month = aug,
       volume = {267},
       number = {2},
          eid = {41},
        pages = {41},
          doi = {10.3847/1538-4365/ace052},
archivePrefix = {arXiv},
       eprint = {2304.03527},
 primaryClass = {astro-ph.GA},
       adsurl = {https://ui.adsabs.harvard.edu/abs/2023ApJS..267...41K}
}

@ARTICLE{lmd+09,
       author = {{Liu}, F.~S. and {Mao}, Shude and {Deng}, Z.~G. and {Xia}, X.~Y. and {Wen}, Z.~L.},
        title = "{Major dry mergers in early-type brightest cluster galaxies}",
      journal = {\mnras},
         year = 2009,
        month = jul,
       volume = {396},
       number = {4},
        pages = {2003-2010},
          doi = {10.1111/j.1365-2966.2009.14907.x},
archivePrefix = {arXiv},
       eprint = {0904.2379},
 primaryClass = {astro-ph.GA},
       adsurl = {https://ui.adsabs.harvard.edu/abs/2009MNRAS.396.2003L}
}

@ARTICLE{mgr+12,
       author = {{Maughan}, B.~J. and {Giles}, P.~A. and {Randall}, S.~W. and {Jones}, C. and {Forman}, W.~R.},
        title = "{Self-similar scaling and evolution in the galaxy cluster X-ray luminosity-temperature relation}",
      journal = {\mnras},
         year = 2012,
        month = apr,
       volume = {421},
       number = {2},
        pages = {1583-1602},
          doi = {10.1111/j.1365-2966.2012.20419.x},
archivePrefix = {arXiv},
       eprint = {1108.1200},
 primaryClass = {astro-ph.CO},
       adsurl = {https://ui.adsabs.harvard.edu/abs/2012MNRAS.421.1583M}
}

@ARTICLE{mg08,
       author = {{Mazzotta}, Pasquale and {Giacintucci}, Simona},
        title = "{Do Radio Core-Halos and Cold Fronts in Non-Major-Merging Clusters Originate from the Same Gas Sloshing?}",
      journal = {\apjl},
         year = 2008,
        month = mar,
       volume = {675},
       number = {1},
        pages = {L9},
          doi = {10.1086/529433},
archivePrefix = {arXiv},
       eprint = {0801.1905},
 primaryClass = {astro-ph},
       adsurl = {https://ui.adsabs.harvard.edu/abs/2008ApJ...675L...9M}
}

@ARTICLE{mgh+08,
       author = {{McIntosh}, Daniel H. and {Guo}, Yicheng and {Hertzberg}, Jen and {Katz}, Neal and {Mo}, H.~J. and {van den Bosch}, Frank C. and {Yang}, Xiaohu},
        title = "{Ongoing assembly of massive galaxies by major merging in large groups and clusters from the SDSS}",
      journal = {\mnras},
         year = 2008,
        month = aug,
       volume = {388},
       number = {4},
        pages = {1537-1556},
          doi = {10.1111/j.1365-2966.2008.13531.x},
archivePrefix = {arXiv},
       eprint = {0710.2157},
 primaryClass = {astro-ph},
       adsurl = {https://ui.adsabs.harvard.edu/abs/2008MNRAS.388.1537M}
}

@ARTICLE{mer+16,
       author = {{McPartland}, Conor and {Ebeling}, Harald and {Roediger}, Elke and {Blumenthal}, Kelly},
        title = "{Jellyfish: the origin and distribution of extreme ram-pressure stripping events in massive galaxy clusters}",
      journal = {\mnras},
         year = 2016,
        month = jan,
       volume = {455},
       number = {3},
        pages = {2994-3008},
          doi = {10.1093/mnras/stv2508},
archivePrefix = {arXiv},
       eprint = {1511.00033},
 primaryClass = {astro-ph.GA},
       adsurl = {https://ui.adsabs.harvard.edu/abs/2016MNRAS.455.2994M}
}

@ARTICLE{mmp+15,
       author = {{Merten}, J. and {Meneghetti}, M. and {Postman}, M. and {Umetsu}, K. and {Zitrin}, A. and {Medezinski}, E. and {Nonino}, M. and {Koekemoer}, A. and {Melchior}, P. and {Gruen}, D. and {Moustakas}, L.~A. and {Bartelmann}, M. and {Host}, O. and {Donahue}, M. and {Coe}, D. and {Molino}, A. and {Jouvel}, S. and {Monna}, A. and {Seitz}, S. and {Czakon}, N. and {Lemze}, D. and {Sayers}, J. and {Balestra}, I. and {Rosati}, P. and {Ben{\'\i}tez}, N. and {Biviano}, A. and {Bouwens}, R. and {Bradley}, L. and {Broadhurst}, T. and {Carrasco}, M. and {Ford}, H. and {Grillo}, C. and {Infante}, L. and {Kelson}, D. and {Lahav}, O. and {Massey}, R. and {Moustakas}, J. and {Rasia}, E. and {Rhodes}, J. and {Vega}, J. and {Zheng}, W.},
        title = "{CLASH: The Concentration-Mass Relation of Galaxy Clusters}",
      journal = {\apj},
         year = 2015,
        month = jun,
       volume = {806},
       number = {1},
          eid = {4},
        pages = {4},
          doi = {10.1088/0004-637X/806/1/4},
archivePrefix = {arXiv},
       eprint = {1404.1376},
 primaryClass = {astro-ph.CO},
       adsurl = {https://ui.adsabs.harvard.edu/abs/2015ApJ...806....4M}
}

@ARTICLE{mfg93,
       author = {{Mohr}, Joseph J. and {Fabricant}, Daniel G. and {Geller}, Margaret J.},
        title = "{An X-Ray Method for Detecting Substructure in Galaxy Clusters: Application to Perseus, A2256, Centaurus, Coma, and Sersic 40/6}",
      journal = {\apj},
         year = 1993,
        month = aug,
       volume = {413},
        pages = {492},
          doi = {10.1086/173019},
       adsurl = {https://ui.adsabs.harvard.edu/abs/1993ApJ...413..492M}
}

@ARTICLE{mrp+23,
       author = {{Montenegro-Taborda}, Daniel and {Rodriguez-Gomez}, Vicente and {Pillepich}, Annalisa and {Avila-Reese}, Vladimir and {Sales}, Laura V. and {Rodr{\'\i}guez-Puebla}, Aldo and {Hernquist}, Lars},
        title = "{The growth of brightest cluster galaxies in the TNG300 simulation: dissecting the contributions from mergers and in situ star formation}",
      journal = {\mnras},
         year = 2023,
        month = may,
       volume = {521},
       number = {1},
        pages = {800-817},
          doi = {10.1093/mnras/stad586},
archivePrefix = {arXiv},
       eprint = {2302.10943},
 primaryClass = {astro-ph.GA},
       adsurl = {https://ui.adsabs.harvard.edu/abs/2023MNRAS.521..800M}
}

@ARTICLE{mrd+20,
       author = {{Morell}, D.~F. and {Ribeiro}, A.~L.~B. and {de Carvalho}, R.~R. and {Rembold}, S.~B. and {Lopes}, P.~A.~A. and {Costa}, A.~P.},
        title = "{Classification and evolution of galaxies according to the dynamical state of host clusters and galaxy luminosities}",
      journal = {\mnras},
         year = 2020,
        month = may,
       volume = {494},
       number = {3},
        pages = {3317-3327},
          doi = {10.1093/mnras/staa881},
archivePrefix = {arXiv},
       eprint = {2003.13836},
 primaryClass = {astro-ph.GA},
       adsurl = {https://ui.adsabs.harvard.edu/abs/2020MNRAS.494.3317M}
}

@ARTICLE{obp+08,
       author = {{O'Dea}, Christopher P. and {Baum}, Stefi A. and {Privon}, George and {Noel-Storr}, Jacob and {Quillen}, Alice C. and {Zufelt}, Nicholas and {Park}, Jaehong and {Edge}, Alastair and {Russell}, Helen and {Fabian}, Andrew C. and {Donahue}, Megan and {Sarazin}, Craig L. and {McNamara}, Brian and {Bregman}, Joel N. and {Egami}, Eiichi},
        title = "{An Infrared Survey of Brightest Cluster Galaxies. II. Why are Some Brightest Cluster Galaxies Forming Stars?}",
      journal = {\apj},
         year = 2008,
        month = jul,
       volume = {681},
       number = {2},
        pages = {1035-1045},
          doi = {10.1086/588212},
archivePrefix = {arXiv},
       eprint = {0803.1772},
 primaryClass = {astro-ph},
       adsurl = {https://ui.adsabs.harvard.edu/abs/2008ApJ...681.1035O}
}

@ARTICLE{ocn+12,
       author = {{Owers}, Matt S. and {Couch}, Warrick J. and {Nulsen}, Paul E.~J. and {Randall}, Scott W.},
        title = "{Shocking Tails in the Major Merger Abell 2744}",
      journal = {\apjl},
         year = 2012,
        month = may,
       volume = {750},
       number = {1},
          eid = {L23},
        pages = {L23},
          doi = {10.1088/2041-8205/750/1/L23},
archivePrefix = {arXiv},
       eprint = {1204.1052},
 primaryClass = {astro-ph.CO},
       adsurl = {https://ui.adsabs.harvard.edu/abs/2012ApJ...750L..23O}
}

@BOOK{pee93,
       author = {{Peebles}, P.~J.~E.},
        title = "{Principles of Physical Cosmology}",
         year = 1993,
    publisher = "Princeton: Princeton University Press",
          doi = {10.1515/9780691206721},
       adsurl = {https://ui.adsabs.harvard.edu/abs/1993ppc..book.....P}
}

@ARTICLE{phi+02,
       author = {{Peng}, Chien Y. and {Ho}, Luis C. and {Impey}, Chris D. and {Rix}, Hans-Walter},
        title = "{Detailed Structural Decomposition of Galaxy Images}",
      journal = {\aj},
         year = 2002,
        month = jul,
       volume = {124},
       number = {1},
        pages = {266-293},
          doi = {10.1086/340952},
archivePrefix = {arXiv},
       eprint = {astro-ph/0204182},
 primaryClass = {astro-ph},
       adsurl = {https://ui.adsabs.harvard.edu/abs/2002AJ....124..266P}
}

@ARTICLE{pfb+06,
       author = {{Poole}, Gregory B. and {Fardal}, Mark A. and {Babul}, Arif and {McCarthy}, Ian G. and {Quinn}, Thomas and {Wadsley}, James},
        title = "{The impact of mergers on relaxed X-ray clusters - I. Dynamical evolution and emergent transient structures}",
      journal = {\mnras},
         year = 2006,
        month = dec,
       volume = {373},
       number = {3},
        pages = {881-905},
          doi = {10.1111/j.1365-2966.2006.10916.x},
archivePrefix = {arXiv},
       eprint = {astro-ph/0608560},
 primaryClass = {astro-ph},
       adsurl = {https://ui.adsabs.harvard.edu/abs/2006MNRAS.373..881P}
}

@ARTICLE{ree+12,
       author = {{Rawle}, T.~D. and {Edge}, A.~C. and {Egami}, E. and {Rex}, M. and {Smith}, G.~P. and {Altieri}, B. and {Fiedler}, A. and {Haines}, C.~P. and {Pereira}, M.~J. and {P{\'e}rez-Gonz{\'a}lez}, P.~G. and {Portouw}, J. and {Valtchanov}, I. and {Walth}, G. and {van der Werf}, P.~P. and {Zemcov}, M.},
        title = "{The Relation between Cool Cluster Cores and Herschel-detected Star Formation in Brightest Cluster Galaxies}",
      journal = {\apj},
         year = 2012,
        month = mar,
       volume = {747},
       number = {1},
          eid = {29},
        pages = {29},
          doi = {10.1088/0004-637X/747/1/29},
archivePrefix = {arXiv},
       eprint = {1201.1294},
 primaryClass = {astro-ph.CO},
       adsurl = {https://ui.adsabs.harvard.edu/abs/2012ApJ...747...29R}
}

@ARTICLE{rp17,
       author = {{Roberts}, Ian D. and {Parker}, Laura C.},
        title = "{Evidence of pre-processing and a dependence on dynamical state for low-mass satellite galaxies}",
      journal = {\mnras},
         year = 2017,
        month = may,
       volume = {467},
       number = {3},
        pages = {3268-3278},
          doi = {10.1093/mnras/stx317},
archivePrefix = {arXiv},
       eprint = {1702.01782},
 primaryClass = {astro-ph.GA},
       adsurl = {https://ui.adsabs.harvard.edu/abs/2017MNRAS.467.3268R}
}

@ARTICLE{srt+08,
       author = {{Santos}, J.~S. and {Rosati}, P. and {Tozzi}, P. and {B{\"o}hringer}, H. and {Ettori}, S. and {Bignamini}, A.},
        title = "{Searching for cool core clusters at high redshift}",
      journal = {\aap},
         year = 2008,
        month = may,
       volume = {483},
       number = {1},
        pages = {35-47},
          doi = {10.1051/0004-6361:20078815},
archivePrefix = {arXiv},
       eprint = {0802.1445},
 primaryClass = {astro-ph},
       adsurl = {https://ui.adsabs.harvard.edu/abs/2008A&A...483...35S}
}

@ARTICLE{sar86,
       author = {{Sarazin}, Craig L.},
        title = "{X-ray emission from clusters of galaxies}",
      journal = {Reviews of Modern Physics},
         year = 1986,
        month = jan,
       volume = {58},
       number = {1},
        pages = {1-115},
          doi = {10.1103/RevModPhys.58.1},
       adsurl = {https://ui.adsabs.harvard.edu/abs/1986RvMP...58....1S}
}

@ARTICLE{sr19,
       author = {{Soares}, N.~R. and {Rembold}, S.~B.},
        title = "{The dynamic stage of clusters and its influence on the stellar populations of galaxies}",
      journal = {\mnras},
         year = 2019,
        month = mar,
       volume = {483},
       number = {4},
        pages = {4354-4370},
          doi = {10.1093/mnras/sty3356},
archivePrefix = {arXiv},
       eprint = {1812.02499},
 primaryClass = {astro-ph.GA},
       adsurl = {https://ui.adsabs.harvard.edu/abs/2019MNRAS.483.4354S}
}

@ARTICLE{swx+24,
       author = {{Sun}, Hanwen and {Wang}, Tao and {Xu}, Ke and {Daddi}, Emanuele and {Gu}, Qing and {Kodama}, Tadayuki and {Zanella}, Anita and {Elbaz}, David and {Tanaka}, Ichi and {Gobat}, Raphael and {Guo}, Qi and {Han}, Jiaxin and {Lu}, Shiying and {Zhou}, Luwenjia},
        title = "{JWST's First Glimpse of a z > 2 Forming Cluster Reveals a Top-heavy Stellar Mass Function}",
      journal = {\apjl},
         year = 2024,
        month = jun,
       volume = {967},
       number = {2},
          eid = {L34},
        pages = {L34},
          doi = {10.3847/2041-8213/ad4986},
archivePrefix = {arXiv},
       eprint = {2403.05248},
 primaryClass = {astro-ph.GA},
       adsurl = {https://ui.adsabs.harvard.edu/abs/2024ApJ...967L..34S}
}

@ARTICLE{vcn+25,
       author = {{V{\'e}liz Astudillo}, S. and {Carrasco}, E.~R. and {Nilo Castell{\'o}n}, J.~L. and {Zenteno}, A. and {Cuevas}, H.},
        title = "{The effect of dynamical states on galaxy cluster populations: II. Comparison of galaxy properties and fundamental relations}",
      journal = {\aap},
         year = 2025,
        month = oct,
       volume = {702},
          eid = {A251},
        pages = {A251},
          doi = {10.1051/0004-6361/202555199},
archivePrefix = {arXiv},
       eprint = {2504.13337},
 primaryClass = {astro-ph.GA},
       adsurl = {https://ui.adsabs.harvard.edu/abs/2025A&A...702A.251V}
}

@ARTICLE{vkf+06,
       author = {{Vikhlinin}, A. and {Kravtsov}, A. and {Forman}, W. and {Jones}, C. and {Markevitch}, M. and {Murray}, S.~S. and {Van Speybroeck}, L.},
        title = "{Chandra Sample of Nearby Relaxed Galaxy Clusters: Mass, Gas Fraction, and Mass-Temperature Relation}",
      journal = {\apj},
         year = 2006,
        month = apr,
       volume = {640},
       number = {2},
        pages = {691-709},
          doi = {10.1086/500288},
archivePrefix = {arXiv},
       eprint = {astro-ph/0507092},
 primaryClass = {astro-ph},
       adsurl = {https://ui.adsabs.harvard.edu/abs/2006ApJ...640..691V}
}

@ARTICLE{whl09,
       author = {{Wen}, Z.~L. and {Liu}, F.~S. and {Han}, J.~L.},
        title = "{Mergers of Luminous Early-Type Galaxies in the Local Universe and Gravitational Wave Background}",
      journal = {\apj},
         year = 2009,
        month = feb,
       volume = {692},
       number = {1},
        pages = {511-521},
          doi = {10.1088/0004-637X/692/1/511},
archivePrefix = {arXiv},
       eprint = {0810.5200},
 primaryClass = {astro-ph},
       adsurl = {https://ui.adsabs.harvard.edu/abs/2009ApJ...692..511W}
}

@ARTICLE{wh13,
       author = {{Wen}, Z.~L. and {Han}, J.~L.},
        title = "{Substructure and dynamical state of 2092 rich clusters of galaxies derived from photometric data}",
      journal = {\mnras},
         year = 2013,
        month = nov,
       volume = {436},
       number = {1},
        pages = {275-293},
          doi = {10.1093/mnras/stt1581},
archivePrefix = {arXiv},
       eprint = {1307.0568},
 primaryClass = {astro-ph.CO},
       adsurl = {https://ui.adsabs.harvard.edu/abs/2013MNRAS.436..275W}
}

\end{document}